\documentclass[amssymb,amsmath,aps,groupedaddress,reprint,superscriptaddress,twocolumn,showpacs]{revtex4}

\usepackage{graphicx}
\usepackage{dcolumn}
\usepackage{bm}
\usepackage{amsmath}

\begin{document}

\title{Giant optical nonlinearity of graphene in a strong magnetic field}
\author{Xianghan Yao}
\affiliation{Department of Physics and Astronomy, Texas A\&M
University, College Station, TX, 77843 USA}
 \author{Alexey Belyanin}
 \email{belyanin@tamu.edu}
\affiliation{Department of Physics and Astronomy, Texas A\&M
University, College Station, TX, 77843 USA}

\date{20 October 2011}

\begin{abstract}

We present quantum-mechanical density-matrix formalism for calculating the nonlinear optical response of magnetized graphene, valid for arbitrarily strong magnetic and optical fields. We  show that magnetized graphene  possesses by far the highest third-order optical nonlinearity among all known materials. The giant nonlinearity originates from unique electronic properties and selection rules near the Dirac point. As a result, even one monolayer of graphene gives rise to appreciable nonlinear frequency conversion efficiency for incident infrared radiation.

\end{abstract}


\maketitle

Graphene, a two-dimensional monolayer of carbon atoms arranged in a hexagonal lattice, holds many records as far as its mechanical, thermal, electrical, and optical properties are concerned; see. e.g.
\cite{novoselov2011} for the review. With this Letter we would like to add yet another distinction to this list of superlatives: we show that graphene in a strong magnetic field has the highest infrared optical nonlinearity of all materials we know.

Strong optical nonlinearity of graphene, like most of its unique electrical and optical properties, stems from linear dispersion of carriers near the K,K' points of the Brillouin zone. As a result, the electron velocity induced by an incident electromagnetic wave is a nonlinear function of induced electron momentum. Nonlinear electromagnetic response of classical charges with linear energy dispersion has been studied theoretically in \cite{mikhailov2008}. Recently, four-wave mixing in mechanically exfoliated graphene flakes without magnetic field has been observed at near-infrared wavelengths
\cite{hendry2010}. Effective bulk third-order susceptibility was estimated to have a very large value, $\chi^{(3)} \sim 10^{-7}$ esu. This is comparable in magnitude to the resonant intersubband $\chi^{(3)}$ nonlinearity observed in the mid-infrared range for low-doped quantum cascade laser structures
\cite{mosely2004}, which are essentially asymmetric coupled quantum well heterostructures.

Nonlinear cyclotron resonance in graphene was considered theoretically  in \cite{mikhailov2009}, again in the classical limit, by solving the equation of motion ${\bf F} = d{\bf p}/dt$ for a massless charge. Classical approximation can be applied to electrons in low magnetic field that are occupying highly excited Landau levels $n \gg 1$, when energy and momentum quantization are neglected. Here we present rigorous quantum mechanical description of the nonlinear optical response of magnetized graphene, which is valid for arbitrary magnetic fields and electron distributions over Landau levels (LLs). After finding matrix elements of the optical transitions between LLs, we calculate the third-order nonlinear susceptibility using the density-matrix formalism and then evaluate the efficiency of the four-wave mixing process. The magnitude of $\chi^{(3)}$ turns out to be astonishingly large, of the order of $10^{-1}$ esu at mid/far-infrared wavelengths in the field of several Tesla. This leads to a surprisingly high four-wave mixing efficiency of the order of $10^{-4}$ W/W$^3$ per monolayer.

Linear (one-photon) absorption in monolayer and bilayer graphene in arbitrary magnetic fields has been calculated in \cite{abergel2007} using Keldysh Green's function formalism. This approach is inconvenient when it comes to calculating the nonlinear optical response. The density matrix formalism adopted in this paper provides a rigorous, intuitive, and  straightforward framework for calculating the hierarchy of nonlinear optical susceptibilities and  interaction of strong multi-frequency EM fields or ultrashort pulses with graphene. Expressions for one-photon absorbance obtained in \cite{abergel2007} can be retrieved by calculating the linear susceptibility in the limit of a weak monochromatic field.

In the absence of the optical field, the  effective-mass Hamiltonian \cite{Ando:05,Ando:02,Ando:07} for a graphene monolayer (in the xy plane) in the magnetic field $B \hat{z}$, in the vicinity of K and K' points \cite{Wallace} in the nearest-neighbor tight-binding model is given by
\begin{equation}
\hat{H}_0=\upsilon_F\left(\begin{array}{cccc}
0 & \hat{\pi}_x-i\hat{\pi}_y & 0 & 0 \\
\hat{\pi}_x+i\hat{\pi}_y & 0 & 0 & 0 \\
0 & 0 & 0 & \hat{\pi}_x+i\hat{\pi}_y \\
0 & 0 & \hat{\pi}_x-i\hat{\pi}_y & 0
\end{array}\right)
\end{equation}
where $\upsilon_F$ is a band parameter ($10^8$ cm/s) \cite{novoselov2005,zhang2005}, $\hat{\vec{\pi}}=\hat{\vec{p}}+e\vec{A}/c$, $\hat{\vec{p}}$ is the electron momentum operator, and $\vec{A}$ is the vector potential, which is equal to $(0,Bx)$ here.

To simplify notations, we write down the solutions to the Schr\"{o}dinger equation $ \hat{H_0} \Psi =\varepsilon \Psi $
separately  near the K and K' point. For example, near the K point the Hamiltonian is  $\hat{H}_0=\upsilon_F \vec{\sigma} \cdot \vec{\pi}$, where $\vec{\sigma} = (\sigma_x, \sigma_y)$ is a vector of Pauli matrices.
The eigenfunction is specified by two quantum numbers, $n$ and $k_y$, where $n=0,\pm1,\pm2,\cdots$, and $k_y$ is the electron wave vector along y direction \cite{Ando:02}:
\begin{equation}
\Psi_{n,k_y}(r)=\frac{C_n}{\sqrt{L}}\exp(-ik_y y)\left(\begin{array}{c}
{\rm sgn}(n)i^{|n|-1}\phi_{|n|-1}\\
i^{|n|}\phi_{|n|}
\end{array}\right)
\end{equation}
with $ C_n= 1$ when $ n = 0$, $C_n = 1/\sqrt{2}$ when $ n \neq 0$,
and
$$
\phi_{|n|}= \displaystyle \frac{H_{|n|}\left( \displaystyle (x-l_c^2k_y)/l_c\right)}{\sqrt{2^{|n|}|n|!\sqrt{\pi}l_c}}\exp{\left[-\frac{1}{2} \left(\frac{x-l_c^2k_y}{l_c}\right)^2\right]} ,
$$
where $l_c=\sqrt{c\hbar/eB}$ and $H_n(x)$ is the Hermite Polynomial.
The eigen energy is $ \varepsilon_n={\rm sgn}(n)\hbar\omega_c \sqrt{|n|}$, where
$\omega_c =\sqrt{2}\upsilon_F/l_c.$

In the presence of the incident classical optical field $\vec E=(1/2) \hat{e} E_{\omega} e^{-i\omega t}$ polarized along the vector $\hat{e}$ in the x-y plane ($\hat{e}_{LHS}=[\hat{x}-i\hat{y}]/\sqrt{2}$ and $\hat{e}_{RHS}=[\hat{x}+i\hat{y}]/\sqrt{2}$, which denote left-hand and right-hand circularly polarized light), we add the vector potential of incident optical field, $\vec{A}_{opt}=\frac{ic}{\omega}\vec{E}$, to the vector potential of the magnetic field in the generalized momentum operator $\hat{\vec{\pi}}$ in the Hamiltonian. This results in adding the interaction Hamiltonian $\hat{H}_{int}$ to $\hat{H}_0$, where
\begin{equation} \label{int}
\hat{H}_{int} = \upsilon_F \vec{\sigma}\cdot \frac{e}{c}\vec{A}_{opt}
\end{equation}

This linear in $\vec{A}_{opt}$ expression for the interaction Hamiltonian is exact, unlike the case of the kinetic energy operator quadratic in momentum, where the term proportional to $A^2$ is usually neglected. Note also that Eq.~(\ref{int}) does not contain the momentum operator and its matrix elements are simply determined by the matrix elements of $\vec{\sigma}$. This  immediately gives the
selection rules \cite{abergel2007} for the transitions between the LLs:  $\hat{e}_{LHS}$ photons are absorbed when $|n_f|=|n_i|+1$, whereas $\hat{e}_{RHS}$ photons are absorbed when $|n_f|=|n_i| - 1$. Here $n_i$ and $n_f$ indicate initial and final energy quantum numbers of LLs.

Now we can  write a standard time-evolution equation for the density matrix of Dirac electrons in graphene coupled to an arbitrary optical field:
\begin{equation} \label{rho}
\frac{\partial \hat{\rho}}{\partial t}=-\frac{i}{\hbar}[\hat{H_0}+\hat{H}_{int},\hat{\rho}] + \hat{R}(\hat{\rho}).
\end{equation}
Here $\hat{R}(\hat{\rho})$ describes incoherent relaxation due to disorder, interaction with phonons, and many-body carrier-carrier interactions. Equations Eq.~(\ref{rho}) have to be solved together with Maxwell's equations that contain the optical polarization $\vec{P}(\vec{r},t)=(1/V)  {\rm Tr}(\hat{\rho} \cdot \vec{\mu})$ (average dipole moment $\langle \vec{\mu} \rangle$ per unit volume) as a source term. In the perturbative regime, they give rise to the hierarchy of the optical susceptibilities $\chi^{(n)}$ \cite{shen}, but they are also valid for describing non-perturbative coupling to strong fields, interaction with ultrashort pulses etc.

Since graphene is essentially a 2D system, it makes sense to introduce a surface (2D) polarization $P_s$ determined as an average dipole moment per unit area rather than unit volume. Below we will use 2D susceptibilities unless specified otherwise.

For a weak monochromatic field one can retain only the term $\rho^{(1)}_{mn} = (\rho^{(0)}_{nn} - \rho^{(0)}_{mm}) \langle m | \hat{H}_{int}| n \rangle/(\varepsilon_m - \varepsilon_n - \hbar \omega - i \hbar \gamma_{mn})$   linear with respect to the field and take the sum $\sum_{m,n} \rho_{nm} \vec{\mu}_{mn}$ to obtain an expression for the linear susceptibility:
\begin{eqnarray} \label{linear} \displaystyle
\chi^{(1)}(\omega) =\sum\limits_{n\geq 1; \alpha,\alpha'} \displaystyle \frac{2 C_{n-1}^2 e^2 \upsilon_F^2}{\pi l_c^2 \hbar \omega \omega_c (\alpha \sqrt{n} - \alpha' \sqrt{n-1})}\nonumber \\
\times\frac{\left(\nu_{n,\alpha} - \nu_{n-1,\alpha'}\right) }{  (\alpha' \sqrt{n-1} \omega_c -\alpha \sqrt{n} \omega_c - \omega- i \gamma)} .
\end{eqnarray}

Here we used $\langle m | \hat{H}_{int}| n \rangle = - (i/\omega) e \upsilon_F \langle m | \vec{\sigma}| n \rangle \vec{E}(\omega)$ and $\langle m | \vec{\mu}| n \rangle =  (i \hbar/(\varepsilon_n - \varepsilon_m)) e \upsilon_F \langle m | \vec{\sigma}| n \rangle$. Note that the matrix element of the interaction Hamiltonian can be written as $-  \tilde{\vec{\mu}}_{mn} \vec{E}$, where $\tilde{\vec{\mu}}_{mn} =  (i/\omega) e \upsilon_F \langle m | \vec{\sigma}| n \rangle $, and  $\tilde{\vec{\mu}}_{mn} = \vec{\mu}_{mn}$  when $\varepsilon_n - \varepsilon_m = \hbar \omega$.

We assumed for simplicity that the relaxation term for the off-diagonal density matrix elements $R_{mn} = -\gamma_{mn} \rho_{mn}$ and all $\gamma$'s are the same. For easy comparison, we used the same notations for LLs as in \cite{abergel2007}: $\alpha,\alpha' = \pm$ denote whether the corresponding state belongs to the conduction (+) or valence (-) band and $\nu_{n,\alpha}$ are the filling factors of LLs; a complete occupation corresponds to $\nu = 2$. The degeneracy of a given LL is $2/(\pi l_c^2)$ including both spin and valley degeneracy. After calculating the dimensionless linear absorbance as $(2 \pi \omega/c) {\rm Im} [\chi^{(1)}(\omega)]$ we obtain the same result as in  \cite{abergel2007}.

Now we consider a specific example of the nonlinear optical interaction, namely the four-wave mixing.
Consider a strong bichromatic field $\vec{E} = (1/2) (\vec{E}_1\exp(-i\omega_1 t) + \vec{E}_2 \exp(-i\omega_2 t) + {\rm c.c.})$ normally incident on the graphene layer. Here $\omega_1$ is nearly resonant with the transition from $n = -1$ to $n = 2$ and $\vec{E}_1$ has left circular polarization. The frequency $\omega_2$ is nearly resonant with the transition from $n = 0$ to $n = \pm 1$ and $\vec{E}_2$ has linear polarization, so that it couples both to transition $-1 \rightarrow 0$ and $0 \rightarrow 1$, as shown in Fig. 1. As a result of the four-wave mixing interaction, the right-circularly polarized field $\vec{E}_3$ at frequency $\omega_3 = \omega_1 - 2 \omega_2$ nearly resonant with the transition from $n = 2$ to $n = 1$ is generated.

Efficient nonlinear mixing becomes possible due to strong non-equidistancy  of the LLs and unique selection rules $\Delta |n| = \pm 1$ which enable transitions with change in $n$ greater than 1, for example the transition from state $n = -1$ to state $n = 2$. This transition would be forbidden in conventional LL systems with $\Delta n = \pm 1$ selection rule. The effective dipole moments for all transitions shown in Fig. 1b scale as $\upsilon_F/\omega$, i.e. they are similar to each other within a factor of 2 and are very large: of the order of 10-100 $\AA$ in the mid/far-IR range. This, in combination with sharp peaks in the density of states at LLs  enables a strong nonlinear response.

\begin{figure}[htb]
\centerline{
\includegraphics[width=8.6cm]{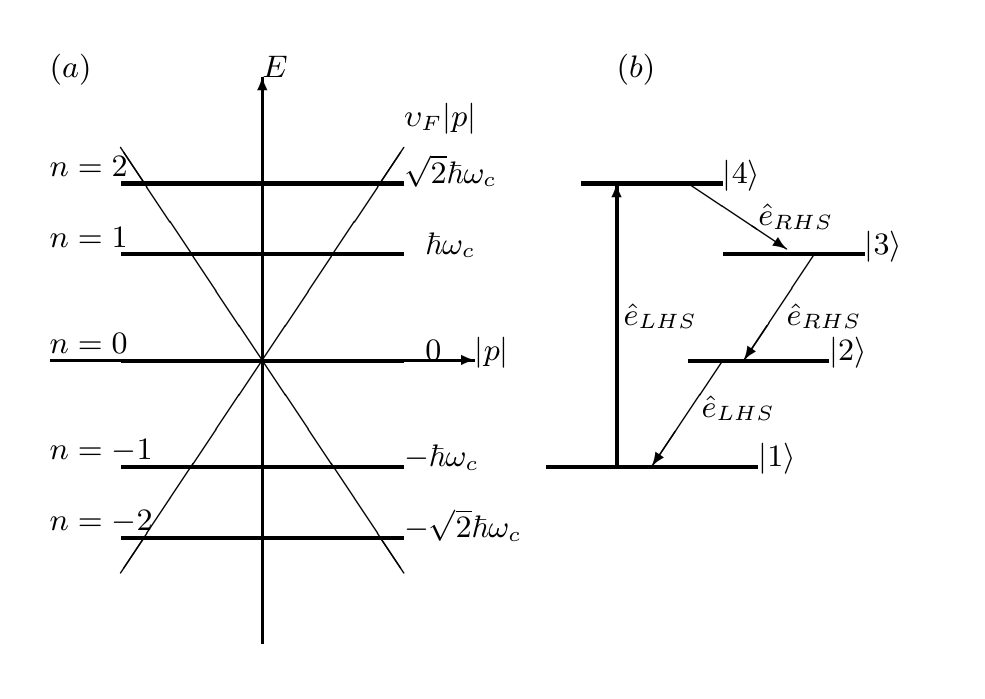}}
  \caption{(a):  Landau levels near the K point superimposed on the electron dispersion without magnetic field $E=\pm\upsilon_F |p|$. (b): A scheme of resonant four-wave mixing process in the four-level system of LLs with energy quantum numbers $n=-1,0,+1,+2$. The case of exact resonance is shown. Polarization of light corresponds to the allowed transitions. }
  \end{figure}

For simplicity we assume that the incident field is not strong enough to significantly modify populations and all states below $ n = 0$ are fully occupied. Then the optical fields interact resonantly only with states $n = -1,0,1,2$, which we renamed to $n = 1,2,3,4$  in Fig. 1b. The Hamiltonian can be truncated to a 4x4 matrix, where $(H_0)_{mn}$ is diagonal, with diagonal elements being the energies of corresponding LLs, and the interaction Hamiltonian is given by the matrix $-  \tilde{\vec{\mu}}_{mn} \vec{E}$ as specified above. This approximation is similar to the one adopted in
\cite{mosely2004,belyanin2003,belyanin2005,belyanin2007} for analyzing resonant nonlinear processes in coupled quantum-well heterostructures.  The resulting third-order nonlinear optical susceptibility at frequency  $\omega_3 = \omega_1 - 2 \omega_2$ is
\begin{eqnarray} \label{chi3}
\chi^{(3)}(\omega_{3})=\frac{-(2/\pi l_c^2) \mu_{43}\tilde{\mu}_{41}\tilde{\mu}_{32}\tilde{\mu}_{21}}{(i\hbar)^3\Gamma_{43}}\nonumber\\
\times\bigg(-\frac{\rho_{33}-\rho_{22}}{\Gamma^*_{31}\Gamma^*_{32}}+\frac{\rho_{22}-\rho_{11}}{\Gamma^*_{31}\Gamma^*_{21}}+\nonumber\\
\frac{\rho_{44}-\rho_{11}}{\Gamma_{42}\Gamma_{41}}+\frac{\rho_{22}-\rho_{11}}{\Gamma_{42}\Gamma^*_{21}}\bigg)\end{eqnarray}
Here the complex detuning factors are $\Gamma_{21} = \gamma_{21} + i((\varepsilon_2 - \varepsilon_1)/\hbar - \omega_2)$,  $\Gamma_{32} = \gamma_{32} + i((\varepsilon_3 - \varepsilon_2)/\hbar - \omega_2)$,  $\Gamma_{41} = \gamma_{41} + i((\varepsilon_4 - \varepsilon_1)/\hbar - \omega_1)$,  and $\Gamma_{43} = \gamma_{43} + i((\varepsilon_4 - \varepsilon_3)/\hbar - \omega_3)$.

In deriving Eq.~({\ref{chi3}) from Eq.~({\ref{rho}) we assumed that populations $\rho_{mm}$ are constant and solved for the off-diagonal density matrix elements. To get an order-of-magnitude estimate, we assume that all fields are in exact resonance and all dephasing rates are the same, so that $\Gamma_{ij} = \gamma$. We also assume for definiteness that state 1 is fully occupied while states 2,3, and 4 are empty. Coming back to original notations of LLs in  Fig. 1a, this means that the $n = 0$ LL  is empty, i.e. the Fermi level is between states $n = 0$ and $n = -1$. The magnitude of $\chi^{(3)}$ will be similar for any distribution of carriers as long as not all population differences in Eq.~({\ref{chi3}) are zero. Then we obtain
\begin{equation}
\label{est}
\chi_{2D}^{(3)} \sim \frac{(6/\pi l_c^2) \mu_{43}\tilde{\mu}_{41}\tilde{\mu}_{32}\tilde{\mu}_{21}}{(\hbar \gamma)^3} \sim 3.6\times 10^{-8} \frac{1}{B(T)} \; {\rm esu},
\end{equation}
where the magnetic field $B(T)$ is expressed in Tesla and we took $\gamma = 2\times 10^{13}$ s$^{-1}$ for the dephasing rates.

This is a 2D (surface) susceptibility. To compare with bulk materials, we divide Eq.~(\ref{est}) by the monolayer thickness $\sim 3$ $\AA$ to obtain the bulk susceptibility $\chi_{3D}^{(3)} \sim 1\, (1/B(T))$ esu. This is by far the strongest nonlinearity as compared to any material that we are aware of. The frequencies involved in the four-wave mixing process fall into the mid/far-IR range for the magnetic field of a few Tesla, as shown in Fig. 2. In particular, at $B = 1$ T the generated nonlinear signal is at the wavelength of about 82 $\mu$m.

The magnitude of the $\chi^{(3)}$ nonlinearity scales roughly as $1/\gamma^3$, i.e. it rapidly decreases with increasing line broadenings.  However, even in a very disordered material with broadenings of the order of transition frequencies $\sim 10^{14}$ s$^{-1}$ the magnitude of $\chi^{(3)}$ is still record-high: above $10^{-5}$ esu.

\begin{figure}[htb]
\centerline{
\includegraphics[width=8.6cm]{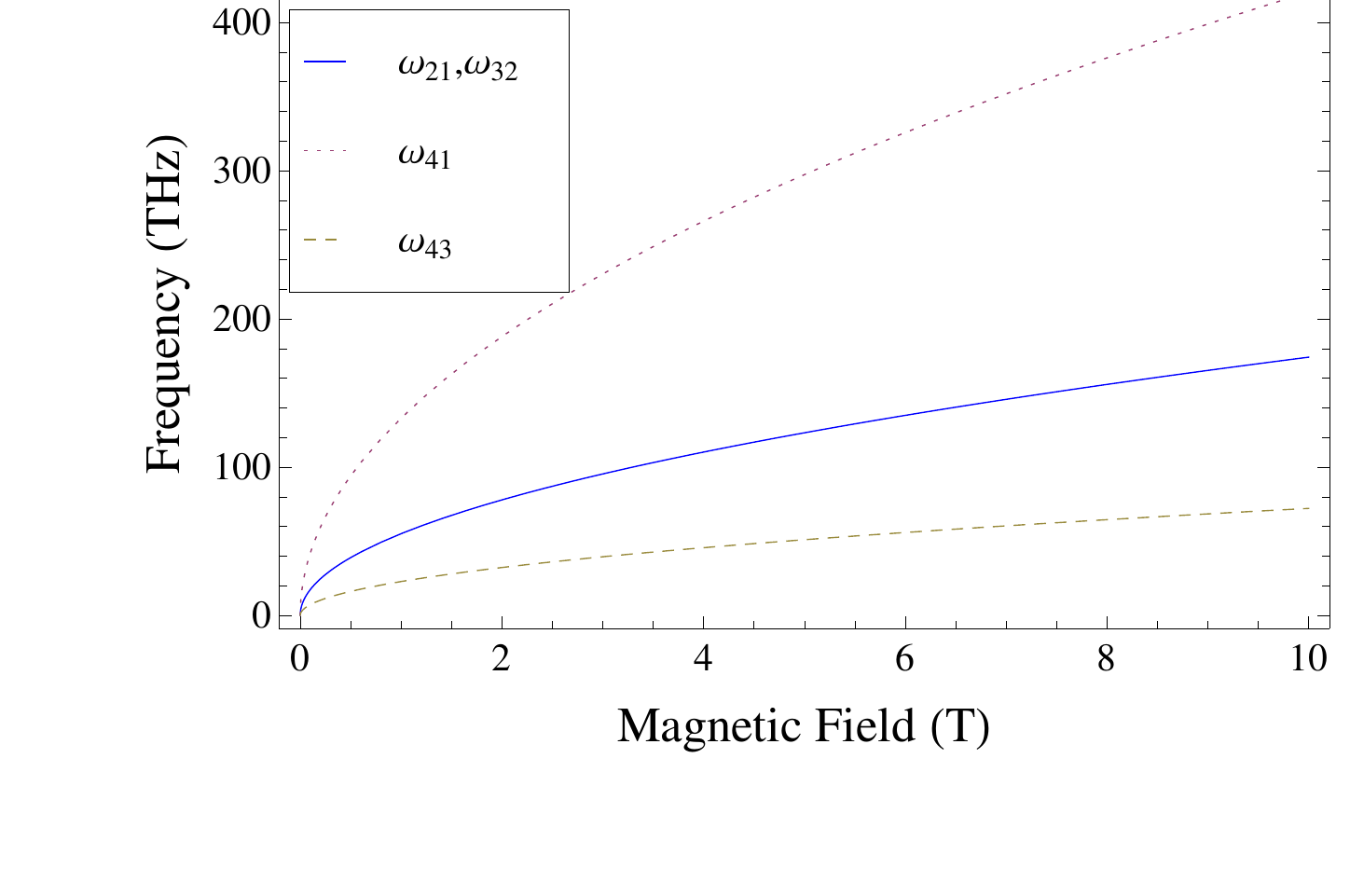}}
   \caption{Transition frequencies in the above 4-energy level graphene system. $\omega_{ij}$ indicates the transition frequency between level $i$ and $j$. }
   \label{fig:transition frequency}
\end{figure}

From wave equations, the electric field amplitude $E_3(\omega_3)$ of the generated nonlinear signal is given by
$ E_3 = i (2\pi\omega_3/c) \chi_{2D}^{(3)}(\omega_3) E_1 E_2^2. $
Assuming that all beams ideally overlap within the area $A$ of the graphene sample, the power of the nonlinear signal from one monolayer is
\begin{equation}
P(\omega_3)= \left( \frac{16 \pi^2 \omega_3}{c^2 A} \right)^2 \left(\chi^{(3)} \right)^2 P_1(\omega_1) (P_2(\omega_2))^2.
\end{equation}

For the illuminated area $A=10^{-4}$ cm$^2$, the power conversion efficiency for the nonlinear signal generation scales  as $\sim 10^{-4} (1/B(T))$ W/W$^3$. This is a remarkably large efficiency for one monolayer of material. It can be further increased by stacking several layers of graphene, e.g. by fabricating non-Bernal stacked epitaxial graphene layers as in recent demonstration of light amplification in graphene \cite{otsuji}.

The above expression for the nonlinear power becomes invalid at very high optical fields, when the effective Rabi frequencies $\tilde{\mu}_{mn} E/\hbar$ of the optical fields become of the order of $\sqrt{\gamma/\tau_{mn}}$, where $\tau_{mn}$ is population relaxation time between states $m$ and $n$. For the magnetic field of several Tesla and $\tau \sim 1$ ps, this corresponds to intensities of about $10^4$ W/cm$^2$. At higher incident fields, population differences in Eq.~(\ref{chi3}) become optically saturated and eventually decrease as $1/E^2$. As a result, the growth of the nonlinear power  with increasing incident power slows down. At even higher optical fields, when the Rabi frequencies exceed the line broadenings $\gamma$, the broadening factors $\Gamma$'s at the transitions driven by strong optical fields increase as  $E^2$ and the nonlinear power decreases with increasing incident power $P_1 \sim P_2 \sim P$ as $1/P$. To analyze the strong-field  case quantitatively, one needs to solve Eq.~({\ref{rho}) for both diagonal and off-diagonal density matrix elements, which is more tedious but straightforward.

In conclusion, graphene in a strong magnetic field possesses record-high optical nonlinearity due to unique properties of quantized Landau levels  near the Dirac point and selection rules for the optical transitions between Landau levels.
High nonlinearity leads to significant nonlinear frequency conversion efficiency even for one monolayer of material. The nonlinearity is expected to be ultrafast, enabling response to THz modulation. These unique properties of magnetized graphene may have important implications for coherent nonlinear generation and detection in the mid-infrared and THz range.

This work was supported in part by NSF Grants ECS-0547019, OISE-0968405, and EEC-0540832.


\begin{thebibliography}{20}

\bibitem[1]{novoselov2011} K.S. Novoselov, Rev. Mod. Phys. 83, 837 (2011).
\bibitem[2]{mikhailov2008} S. A. Mikhailov and K. Ziegler, J. Phys.: Condens. Matter 20, 384204 (2008).
\bibitem[3]{hendry2010} E. Hendry, P.J. Hale, J. Moger, and A.K. Savchenko, S.A. Mikhailov, Phys. Rev. Lett. 105, 097401 (2010).
\bibitem[4]{mosely2004} T.S. Mosely, A. Belyanin, C. Gmachl, D.L. Sivco, M.L. Peabody, and A.Y. Cho, Optics Express 12, 2972 (2004).
\bibitem[5]{mikhailov2009} S. A. Mikhailov, Phys. Rev. B 79, 241309(R) (2009).
\bibitem[6]{abergel2007} D.S.L. Abergel and V. I. Fal'ko, Phys. Rev. B 75, 155430 (2007).
\bibitem[7]{Ando:05} T. Ando, J. Phys. Soc. Jpn. 74, 777 (2005).
\bibitem[8]{Ando:02} Y. Zheng and T. Ando, Phys. Rev. B 65, 245420 (2002).
\bibitem[9]{Ando:07} T. Ando, J. Phys. Soc. Jpn. 76, 024712 (2007).
\bibitem[10]{Wallace} P. R. Wallace, Phys. Rev. 7, 622 (1947).
\bibitem[11]{novoselov2005} K.S. Novoselov, A.K. Geim, S.V. Morozov, D. Jiang, M.I. Katsnelson, I.V. Grigorieva, S.V. Dubonos and A.A. Firsov, Nature 438, 197 (2005).
\bibitem[12]{zhang2005} Y. Zhang Y, Y-W. Tan, H.L. Stormer, and P. Kim, Nature 438, 201 (2005).
\bibitem[13]{shen} Y.R. Shen, {\it The Principles of Nonlinear Optics}, J. Wiley \& Sons, Hoboken (2003).
\bibitem[14]{belyanin2003} C. Gmachl, A. Belyanin, D.L. Sivco, Milton L. Peabody, N. Owschimikow, A. M. Sergent, F. Capasso, and A.Y. Cho, IEEE Journal of Quant. Electron., 39, 1345 (2003).
\bibitem[15]{belyanin2005} M. Troccoli, A. Belyanin, F. Capasso, E. Cubukcu, D. L. Sivco, and A.Y. Cho, Nature, 433, 845 (2005).
\bibitem[16]{belyanin2007} M. Belkin, F. Capasso, A. Belyanin, D. L. Sivco, A. Y. Cho, D. C. Oakley, C. J. Vineis, and G. W. Turner, Nature Photonics, 1, 288 (2007).
\bibitem[17]{otsuji} H. Karasawa, T. Komori, T. Watanabe, A. Satou,
H. Fukidome, M. Suemitsu, V. Ryzhii, and T. Otsuji, J. Infrared, Mill. THz Waves 32, 655 (2011).

\end{thebibliography}
\end{document}